\documentclass[9pt]{article}

\usepackage{spconf,graphicx}
\usepackage{float}
\usepackage{multirow}
\usepackage{amssymb,amsmath,amsthm,enumitem}

\usepackage[belowskip=-10pt,aboveskip=2pt]{caption} 
\ninept 
\newlength\myindent 
\def\Width{0\kern2\tabcolsep\ldots\kern1\tabcolsep0} 
\usepackage{etoolbox} 
\newcommand{\zerodisplayskips}{% 
  \setlength{\abovedisplayskip}{3pt} 
  \setlength{\belowdisplayskip}{3pt} 
  \setlength{\abovedisplayshortskip}{3pt} 
  \setlength{\belowdisplayshortskip}{3pt}} 
\appto{\normalsize}{\zerodisplayskips} 
\appto{\small}{\zerodisplayskips} 
\appto{\footnotesize}{\zerodisplayskips}

\title{Knowledge Transfer for Efficient On-device False Trigger Mitigation}
\name{Pranay Dighe, Erik Marchi, Srikanth Vishnubhotla, Sachin Kajarekar, Devang Naik}
\address{Apple, One Apple Parkway, Cupertino, USA}
\begin{document}
\setlength{\abovedisplayskip}{3pt}
\setlength{\belowdisplayskip}{3pt}

\maketitle

\begin{abstract}
\vspace{-3mm}
In this paper, we address the task of determining whether a given utterance is directed towards a voice-enabled smart-assistant device or not. An undirected utterance is termed as a \textit{``false trigger''} and false trigger mitigation (FTM) is essential for designing a privacy-centric non-intrusive smart assistant. The directedness of an utterance can be identified by running automatic speech recognition (ASR) on it and determining the user intent by analyzing the ASR transcript. But in case of a false trigger, transcribing the audio using ASR itself is strongly undesirable. To alleviate this issue, we propose an LSTM-based FTM architecture which determines the user intent from acoustic features directly without explicitly generating ASR transcripts from the audio. The proposed models are small-footprint and can be run on-device with limited computational resources. During training, the model parameters are optimized using a knowledge transfer approach where a more accurate self-attention graph neural network model \cite{dighe2020gcn} serves as the teacher. Given the whole audio snippets, our approach mitigates 87\% of false triggers at 99\% true positive rate (TPR), and in a streaming audio scenario, the system listens to only 1.69s of the false trigger audio before rejecting it while achieving the same TPR.
\end{abstract}


\vspace{-6mm}
\section{Introduction}
\label{sec:intro}
\vspace{-4mm}
User privacy is a fundamental concern in the design of smart-assistants on devices like mobile phones, smart speakers, watches, etc. Voice-operated smart-assistants typically rely on detecting a trigger-phrase with high fidelity before they send user audio to the server for ASR and further processing \cite{sigtia2020multitask}. However, on rare occasions, the trigger-phrase detection algorithm may be triggered by unintended audio which might be sent to the server for processing. In addition, there are some use-cases where the processing of user audio is not gated by detection of a trigger-phrase, e.g., when the device is listening to the user for a response to a device-generated query. In these cases, the smart assistant relies on other mechanisms than trigger-phrase detection to determine if the user intends to speak to the device or not. One such mechanism uses self-attention based graph neural networks \cite{dighe2020gcn,sperber-etal-2019-self} (called \textit{``LatticeGNN''} hereafter) to extract information from ASR lattices to detect whether the user audio contains spoken content directed towards the device or not. While LatticeGNN ensures that falsely triggered audio is accurately identified, there are limitations in its usage. Firstly, it is computationally heavy to run full-fledged general purpose ASR on devices with low-resource hardware. Second, LatticeGNN approach requires running ASR on full utterances and an early detection of false triggers is not possible. Moreover, any ASR based approach results in transcribing unintended audio, thereby raising privacy concerns. In contrast, an ideal user intent classifier system should be able to detect \textit{unintended} audio as soon as possible without actually transcribing the spoken content in the audio. We address this task in the current work using on-device LSTM based FTM models.

\begin{figure}[H]
  \centering
  \includegraphics[width=0.8\linewidth]{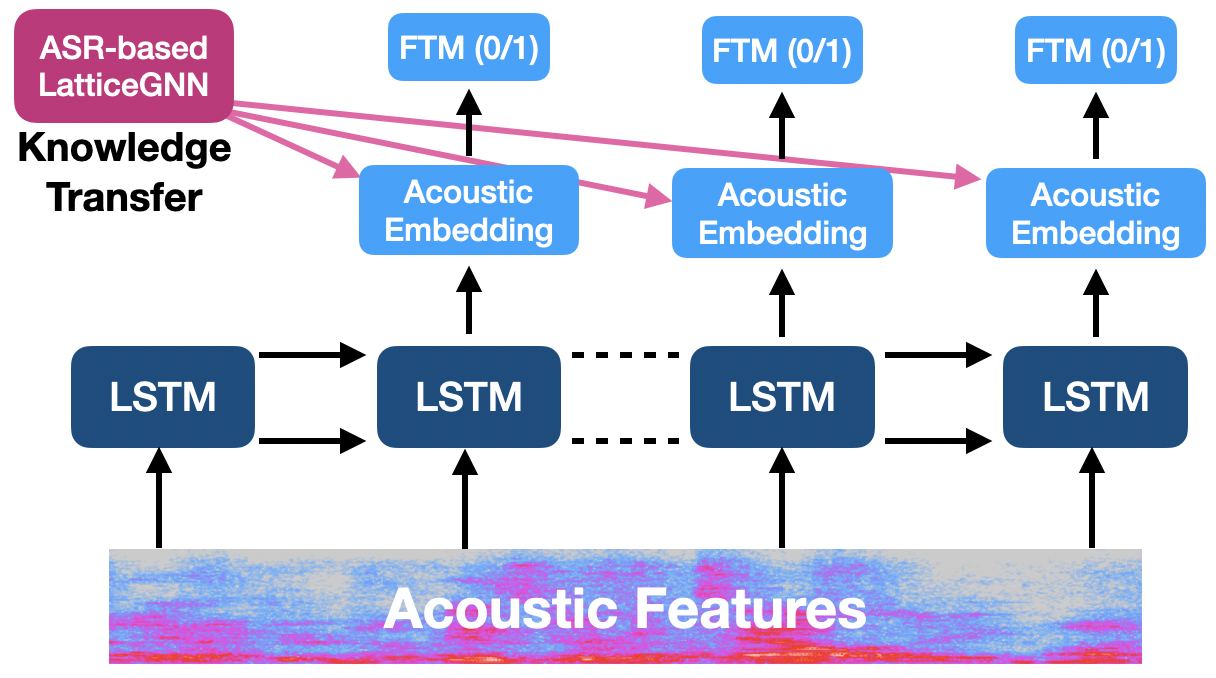}
  \caption{LSTM-based acoustic embeddings for FTM.}
  \label{fig:acoustic_ftm}
\end{figure}
\vspace{-2mm}
Our approach is motivated by the hypothesis that acoustic embeddings may provide extra and complementary information to the ASR based cues for the FTM task. While ASR lattices focus on high-level information such as word/sentence hypotheses for determining device-directedness of the utterance, the acoustic embeddings focus on low-level information such as presence of speech, background noise, and acoustic environment. By using knowledge transfer, we aim to combine these complementary sources of information in a common acoustic embedding based model. 

Utterance-level acoustic embeddings computed using LSTMs have been previously used in \cite{maas2018combining,mallidi2018device,haung2019study} where they were combined with ASR decoding features for device-directed audio detection. Our approach depicted in Figure \ref{fig:acoustic_ftm} differs from prior literature in multiple ways though. Firstly, we propose to make frame level decisions for device-directedness detection so that a falsely triggered audio can be detected and suppressed as soon as possible in a streaming audio scenario. In comparison, the prior literature makes FTM decisions only when the whole utterance has been processed which leads to unnecessary processing of unintended audio. Second, we perform knowledge transfer from an ASR-based LatticeGNN model during training of our LSTM models in order to achieve at least as good accuracy as the teacher LatticeGNN model. The trained LSTM models which generate acoustic embeddings have a small footprint so that they can run on the device. Third, our approach does not need to run ASR during inference and we do not perform model combination before making the final decision to mitigate a false trigger. So, the model is ASR-free at inference time. Further details on knowledge transfer training and streaming FTM decisions are presented in Section \ref{sec:approach}. Other prior approaches for device-directed utterance detection includes various trigger-phrase detection techniques explored in \cite{sigtia2018vt,kumatani2017direct,wu2018monophone,guo2018time,norouzian2019exploring}. Lattice-based techniques which complement trigger-phrase detection systems have been explored in \cite{jeon2019latrnn,haung2019study,dighe2020gcn}.

The rest of the paper is organized as follows. Section \ref{sec:approach} provides a brief background on LatticeGNN and the knowledge transfer approach to train LSTM-FTM models. Section \ref{sec:experiments} provides details of streaming on-device FTM experiments, results and analysis. Finally, Section \ref{sec:conclusions} draws conclusions of this work.

\vspace{-2mm}
\section{Our Approach}
\label{sec:approach}
\vspace{-2mm}
This section provides details of various components involved in the knowledge transfer approach for streaming FTM.

\subsection{LatticeGNN FTM and Lattice Embeddings}
\label{sec:LatticeGNN}
\vspace{-1mm}
LatticeGNN is a graph neural network that uses self-attention mechanism to process word lattices produced by ASR \cite{dighe2020gcn}. This approach can be motivated by the hypothesis that a device-directed utterance has speech content related to the device functions. Such utterances are often less noisy and the best sentence hypothesis has zero (or few) competing hypotheses in the ASR lattice. In contrast, false triggers typically originate either from background noise or from speech which sounds similar to the trigger-phrase. Multiple ASR hypotheses may be present as alternate paths in the lattices of false trigger utterances. These properties of the lattices can be analyzed using the LatticeGNN.
%LatticeGNN technique for FTM can be motivated by the hypothesis that a device-directed utterance has speech content related to the device functions. Such utterances are often less noisy and the best sentence hypothesis has zero (or few) competing hypotheses in the ASR lattice. In contrast, false triggers typically originate either from background noise or from speech which sounds similar to the trigger-phrase. Multiple ASR hypotheses may be present as alternate paths in the lattices of false trigger utterances. Such properties of lattices can be analyzed using a LatticeGNN which is a graph neural network that uses self-attention mechanism to process lattice inputs. 
\begin{figure}[H]
  \centering
  \includegraphics[width=0.6\linewidth]{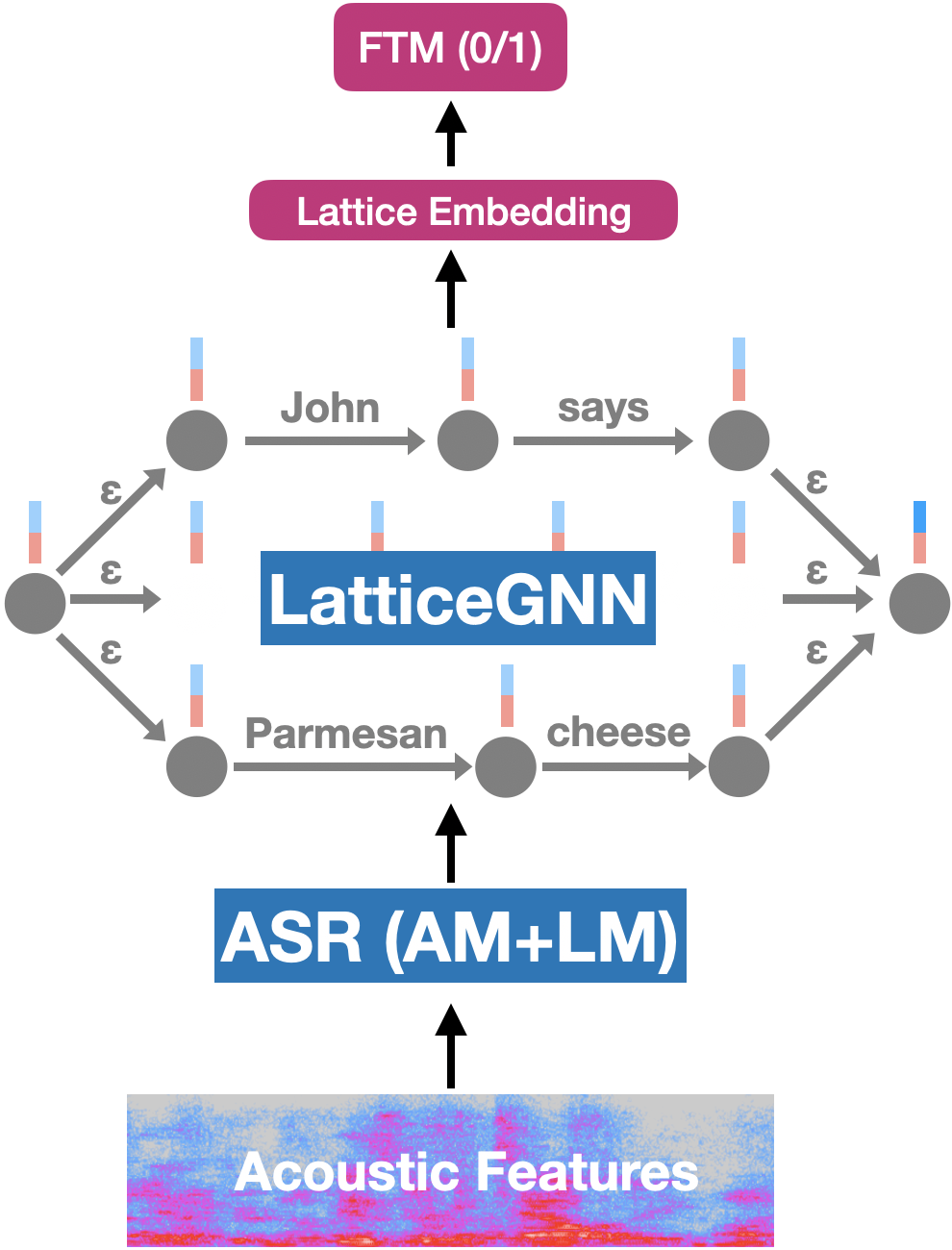}
  \caption{LatticeGNN summarizes ASR lattice information in a lattice embedding.}
  \label{fig:LatticeGNN}
\end{figure}
Figure \ref{fig:LatticeGNN} depicts the LatticeGNN approach. LatticeGNN summarizes the lattice information into an embedding vector which is then used to make a decision on the utterance. Lattice embeddings are obtained by treating the lattice as a graph and processing it using multiple hidden layers of multi-headed self-attention operation. These embeddings have been shown to be highly informative for FTM task \cite{jeon2019latrnn,dighe2020gcn}, but they can be obtained only by running full-fledged ASR on the audio which is expensive to be run on-device and invades user privacy in case of a false trigger. Moreover, the LatticeGNN model needs to be retrained if the distribution of the input lattice features changes due to any changes in the acoustic model, language model or the ASR decoding parameters. To alleviate all these concerns, we propose to bypass ASR and estimate the lattice embeddings directly from the acoustic features of the utterances. We use existing knowledge transfer mechanism \cite{hinton2015distilling} as explained below.

\subsection{Knowledge Transfer from LatticeGNN Embeddings to Acoustic Embeddings}
\label{sec:knowledge_transfer}
\begin{figure}[t]
  \centering
  \includegraphics[width=\linewidth]{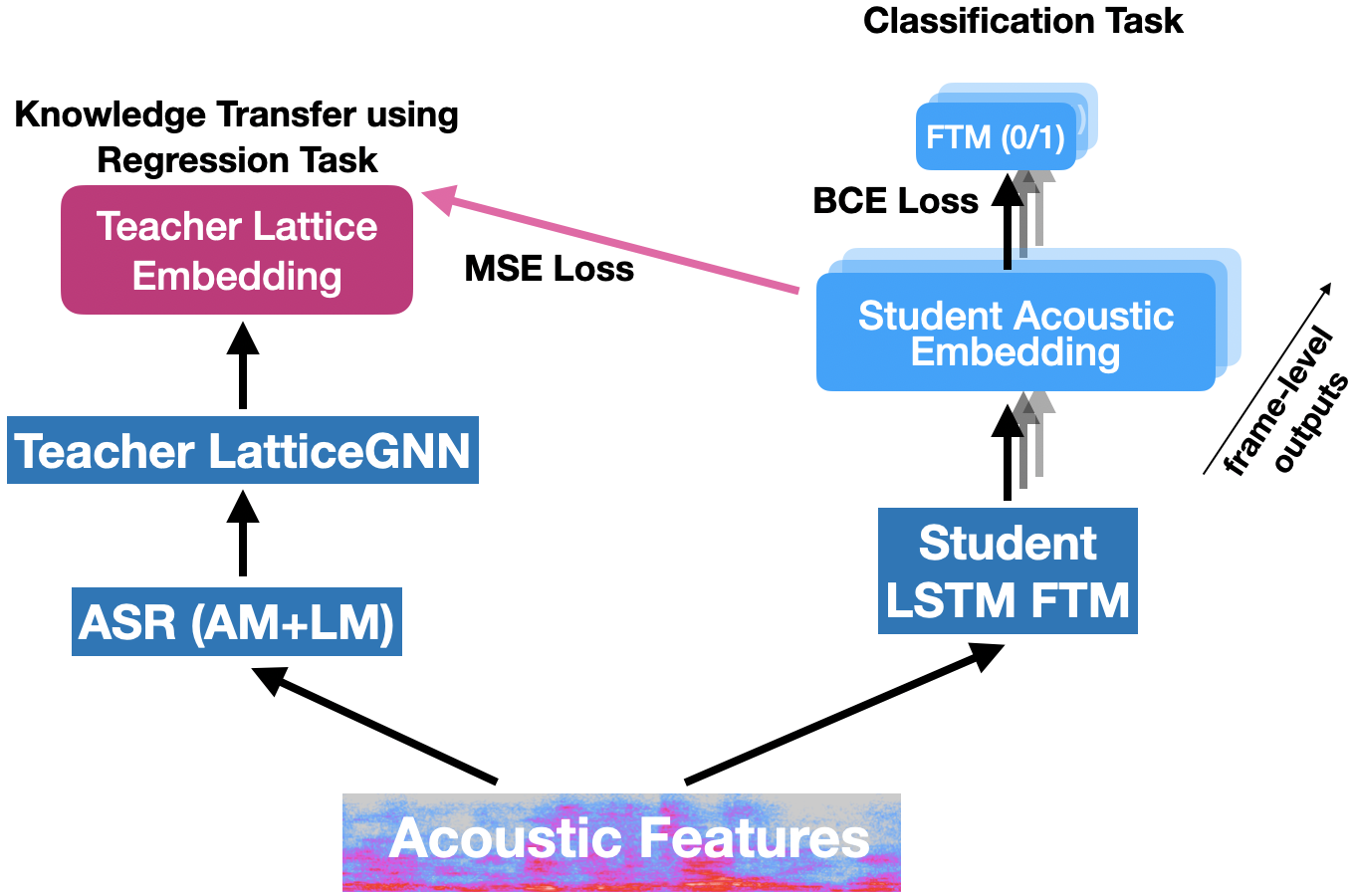}
  \caption{Knowledge transfer approach with LatticeGNN as teacher and on-device LSTM as student model.}
  \label{fig:knowledge_trasnfer}
\end{figure}

\vspace{-1mm}
Figure \ref{fig:knowledge_trasnfer} depicts the knowledge-transfer approach. We start with a pretrained teacher LatticeGNN model which can generate useful lattice embeddings for any given lattice. This teacher model is used to generate lattice embeddings for all the utterances in the training dataset. A student LSTM model is then trained to take a sequence of acoustic features as input and estimate a sequence of frame-level embeddings of the same dimension as the lattice embedding. The acoustic embeddings from the student model are used for two tasks simultaneously: 1) a regression task to match the lattice embedding from the teacher LatticeGNN model and 2) a classification task to determine device-directedness of the utterance. The regression task acts as the bridge for knowledge transfer from the teacher LatticeGNN to the student LSTM model. 

The FTM task is a binary classification between true triggers versus false triggers. The relevant frames of the audio are all labeled as either $1$ for true triggers and $0$  for false triggers. We minimize a binary cross entropy loss $\ell_{BCE}$ computed at each frame of the audio for training this task. We define the \textit{``trigger-phrase detection''} event as the time instant when the beginning of the trigger phrase is detected by the on-device voice-trigger detection mechanism. We do not assign any labels to the first 50 frames (0.5s) of the audio as it contains extra padded audio from before this trigger-phrase detection event. We also assume that enough evidence is not accumulated in first 0.5s to make any predictions. Nevertheless, the first 50 frames are processed by the LSTM for making predictions for the later frames. 

The teacher LatticeGNN model generates a single utterance-level embedding vector whereas the student LSTM generates an acoustic embedding for each frame of the audio. During training, we use the same utterance-level LatticeGNN embedding as the training target for each frame of the audio using a mean squared error loss $\ell_{MSE}$. The overall training objective is to minimize a combined loss defined as:
\begin{equation}
\ell_{combined}=\ell_{BCE}+\alpha\text{ }\ell_{MSE}
\end{equation}
where $\alpha$ is a tunable parameter which is used to weigh the importance of the two tasks during training. After training the student model, we only use the classification task for FTM. The final model is small-footprint and ASR-free.

\subsection{Streaming FTM Using Acoustic Embeddings}
\label{sec:streaming}
\begin{figure}[t]
  \centering
  \includegraphics[width=0.8\linewidth]{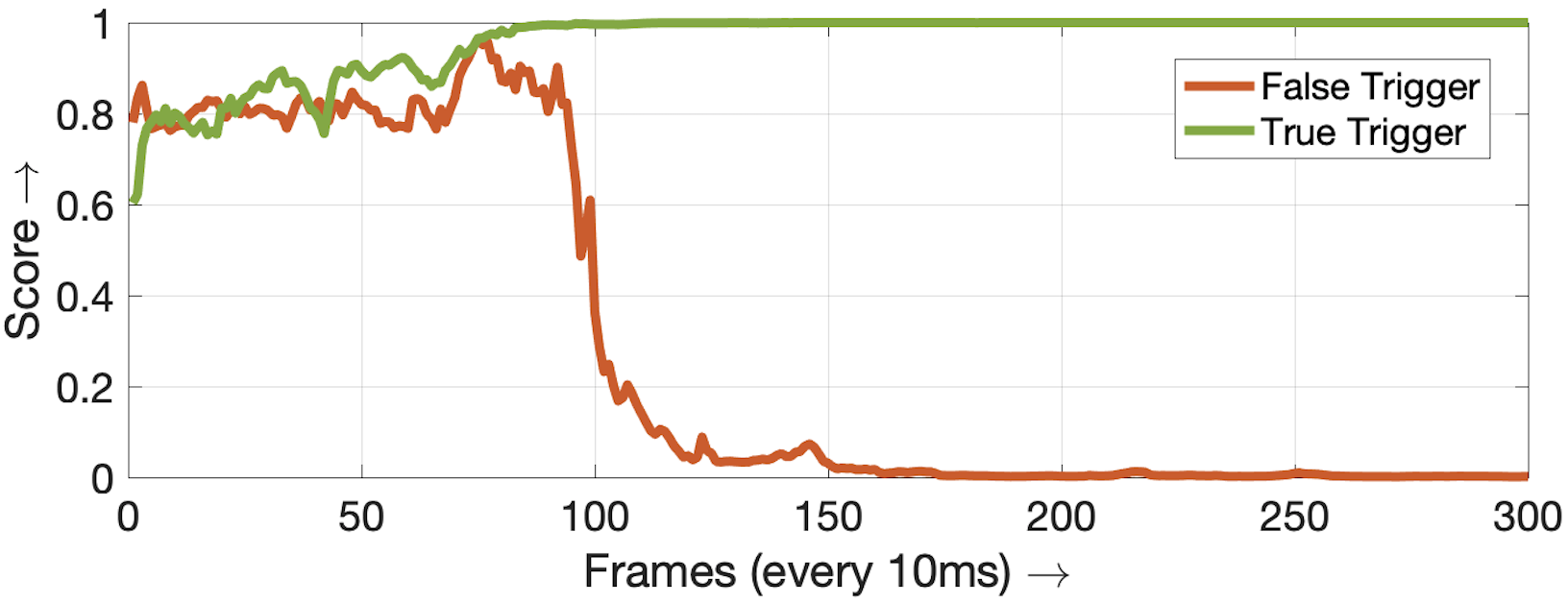}
  \caption{Evolution of mitigation signal over time for an example of a true trigger and a false trigger.}
  \label{fig:mitsignal}
\end{figure}

\vspace{-1mm}
A streaming FTM model can process audio as it arrives and generate an FTM score at each frame. The stream of scores forms a mitigation signal as shown in Figure \ref{fig:mitsignal}. The signal oscillates initially before converging towards 0 or 1 as sufficient audio is processed. A streaming mitigation signal can be used in two ways in practice, as described in the following two sections.

\subsubsection{Fixed Mitigation Delay Time}\label{sec:fmdt}
Mitigation delay time (MDT) $t_d$ is defined as the time delay in rejecting a false alarm. This time is measured from the trigger-phrase detection event at $0.5s$. 
In this approach, we set a fixed mitigation delay time of $t_d$ seconds at which we make the decision to either continue processing the utterance or reject it if its mitigation signal goes below the threshold $\tau$. By defining false alarm rate (FAR) as the fraction of total false triggers that are not rejected and true positive rate (TPR) as the fraction of true triggers that are not rejected, we can tune the threshold $\tau$ between 0 and 1 to obtain a desired FTM performance in terms of low FAR and high TPR. 
%The mitigation signal is smoothed by taking mean of the signal values from the last 50 frames until the frame at $t_d$s.
FTM accuracy should improve as we increase $t_d$ because it allows the LSTM model to accumulate more information about the nature of the utterance. Setting $t_d=$\textit{utterance-length} is equivalent to using the last frame output of the LSTM network and this setting is expected to give the most accurate FTM performance.
\begin{figure}[b]
  \centering
  \includegraphics[width=\linewidth]{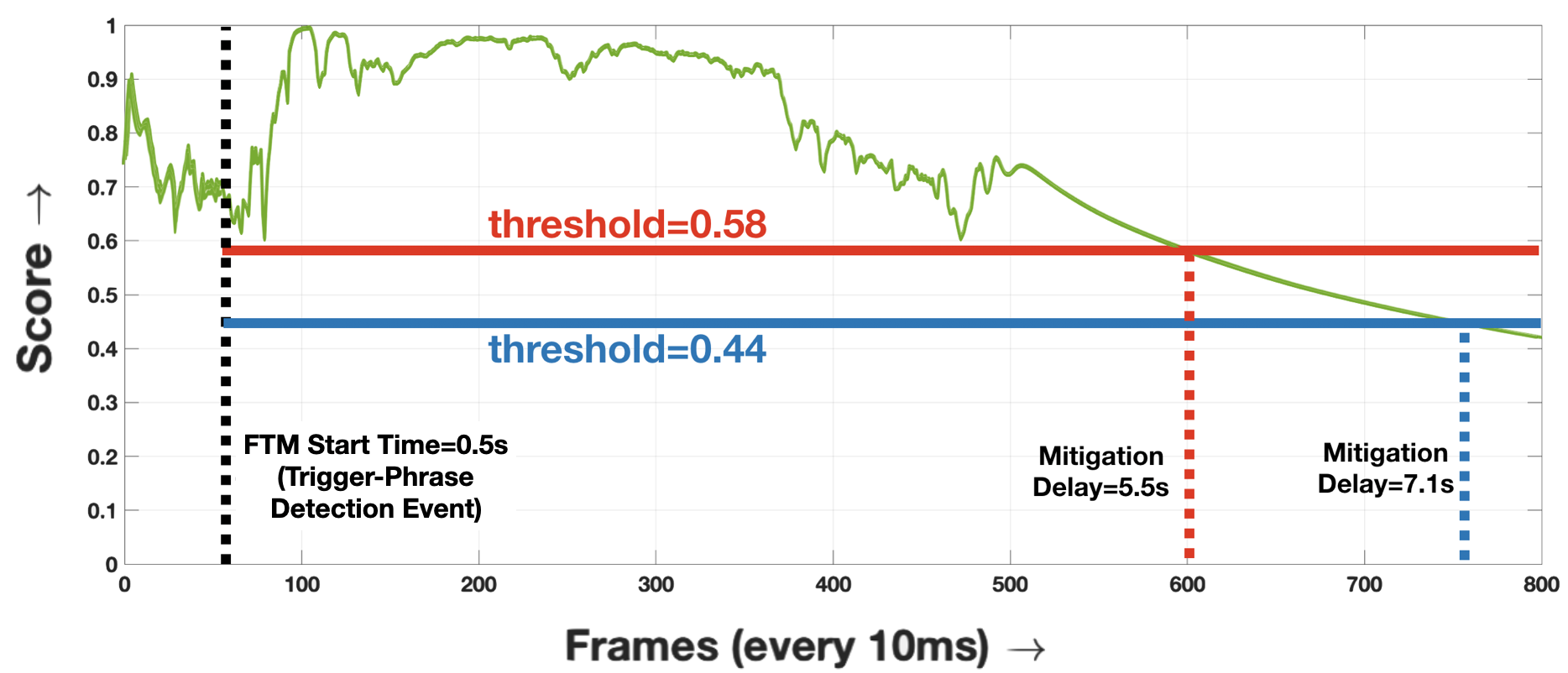}
  \caption{Variable Mitigation Delay Time}
  \label{fig:approach1}
\end{figure}

\subsubsection{Variable Mitigation Delay Time}\label{sec:vmdt}
\vspace{-1mm}
In this approach, each utterance is considered to be a true trigger by default. We start monitoring the mitigation signal from $0.5s$ onward as shown in Figure \ref{fig:approach1}. We fix a threshold $\tau$, (say $\tau=0.58$ for example, depicted by red-line). As long as the mitigation signal stays above $\tau$, we process the audio and keep treating it as device-directed. If the mitigation signal drops below $\tau$ and the utterance hasn't finished yet, we decide that the utterance is a false trigger and stop processing the audio any further. MDT $t_d$ is the delay in mitigating a false trigger under this approach. A short MDT is desired for more privacy. As depicted in Figure \ref{fig:approach1}, the utterance is rejected at a $t_d=5.50s$ after the trigger was detected. The other example depicts the case for $\tau=0.44$ (blue-line). In this case, the utterance is rejected at $t_d=7.10s$. A shorter MDT and higher TPR is desirable. Under this approach, the FTM task will have following properties: $\tau=0.0$ implies $t_d=$\textit{utterance-length} and TPR=100\%; and $\tau=1.0$ implies $t_d=0.0s$, and TPR=0\%.

\section{Experiments and Results}
\label{sec:experiments}
\vspace{-1mm}
In this section, we provide details of our experiments, their results and subsequent analysis.
\vspace{-1mm}
\subsection{Dataset}
\label{sec:data}
\vspace{-1mm}
Our FTM dataset for training the teacher LatticeGNN model consists of device-directed true trigger utterances and unintended false trigger utterances as summarized in Table \ref{tab:dataset}. The model is trained and tuned using \textit{Training} and \textit{CV} partitions respectively. During training of the student LSTM model, we augment the FTM \textit{Training} partition with an additional $\sim1$ million utterances of the true trigger class. These additional utterances provide us with extra training samples for knowledge transfer from the teacher model to the student model. We augment only the true trigger class with additional utterances because true trigger utterances are more readily available to us as compared to much rarer false trigger utterances. Lattice embeddings for the augmented dataset are generated by running ASR and using LatticeGNN on the ASR lattices. These embeddings serve as the training targets for the student model. All the evaluations are done on \textit{Eval} set.
\begin{table}[b]
\small
\centering
\begin{tabular}{lccc}
\hline 
Class & \textit{Training} & \textit{CV} & \textit{Eval} \\ 
\hline 
True Triggers & 133,575 &  14,840  & 13,000 \\
False Triggers & 29,231 & 3,314 & 7,459 \\
\hline 
\end{tabular} 
\caption{FTM dataset.}
\label{tab:dataset}
\end{table}
\vspace{-1mm}
\subsection{Models}
\label{sec:models}
\vspace{-1mm}
The LatticeGNN model is a 2-layer 4-headed masked self-attention based graph neural network with 64-dimensional hidden state and it produces 64-dimensional lattice embeddings. More details on LatticeGNN architecture can be found in \cite{dighe2020gcn}. We consider the combination of LatticeGNN model (with 39,105 parameters) and the underlying ASR system (AM+LM with $\sim60$ million parameters) as the overall teacher model here. The student LSTM model has 4 layers of uni-directional LSTM cells with a hidden dimension of 128. Input features for this model are 40-dimensional log filterbank energy vectors which are computed every 10ms. The output of the last LSTM layer is compressed to a 64-dimensional acoustic embedding at each frame. The student model has $\sim0.5$ million parameters which is $\sim120$x less than the teacher model.

\begin{figure*}[tp]
  \centering
  \includegraphics[width=0.9\linewidth]{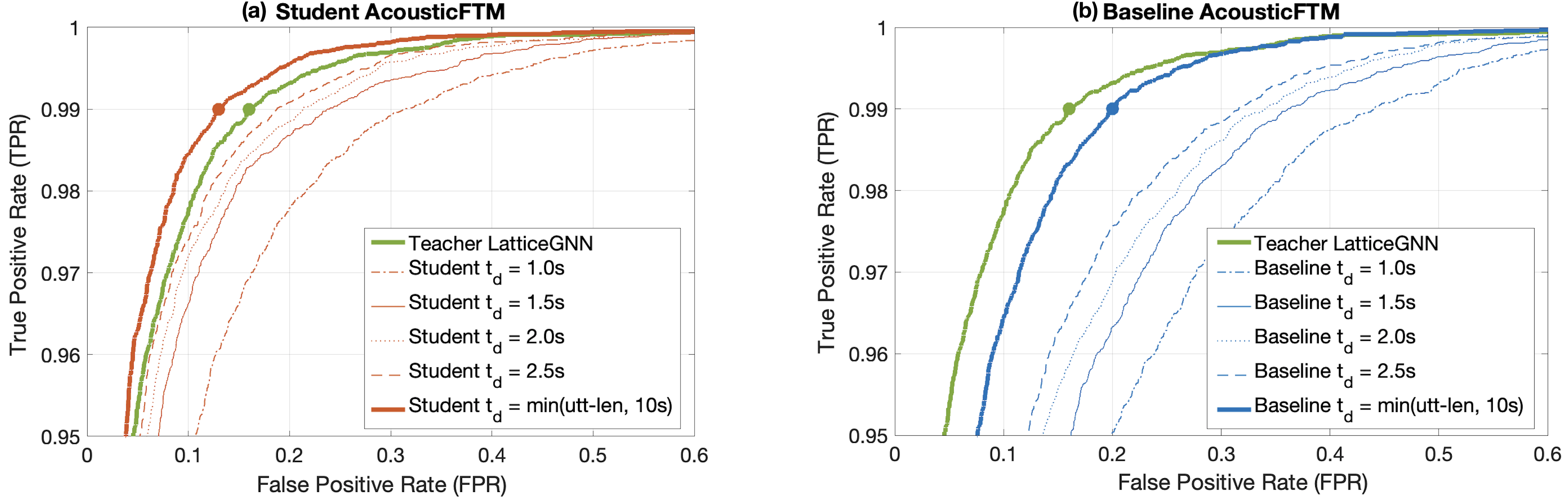}
  \caption{Streaming FTM under fixed MDT approach using (a) student acoustic FTM and (b) baseline acoustic FTM model for different value of $t_d$'s. Teacher LatticeGNN model always processes the whole utterance.}
  \label{fig:fixed_mdt}
\end{figure*}

%\begin{table}[h]
%\small
%\centering
%\begin{tabular}{ccccc}
%\hline 
%\multicolumn{2}{c}{Partition} & LatticeGNN & Additional & Student LSTM \\ 
%& & Training Data & Data & Training Data (Total)\\ 
%\hline 
%\multirow{3}{*}{True} & train & 42,675 & 800,000 & 11,646 \\
%& cv & 4742 & 199460 & 11,646 \\
%& train & 42,675 & 4,742 & 11,646 \\
%\multirow{3}{*}{False} & train & 42,675 & 4,742 & 11,646 \\
%& cv & 42,675 & 4,742 & 11,646 \\
%& eval & 42,675 & 4,742 & 11,646 \\
%\hline 
%\end{tabular} 
%\caption{Dataset for FTM task.}
%\label{tab:dataset}
%\end{table}

%\begin{table}[h]
%\small
%\centering
%i\begin{tabular}{lp{20pt}p{20pt}p{20pt}p{20pt}p{20pt}p{20pt}}
%\hline
%\multirow{2}{*}{Dataset} & \multicolumn{3}{c}{True Triggers} & \multicolumn{3}{c}{False Triggers}\\
%& tr & cv & ev & tr & cv & ev \\
%\hline
%LatticeGNN & 42,675 & 4742 & 6000 & 20000 & 2000 & 6000 \\
%Additional & 800000 & 199460 & - & - & - & - \\
%Total & 842675 & 204202 & 6000 & 20000 & 2000 & 6000 \\
%\hline
%\end{tabular}
%\caption{Dataset for FTM task.}
%\label{tab:dataset}
%\end{table}
\vspace{-2mm}
\subsection{Evaluation Metrics}
\label{sec:metrics}
\vspace{-1mm}
For the fixed MDT approach, we use ROC curves comparing TPR and FAR metrics for different values of MDT. We prefer ROC curves with large area under the curve (AUC) and focus only on a high TPR ($>0.95$) regime in our analysis plots. For variable MDT approach, we define a metric \textit{average mitigation delay time} $\tilde{t_d}$ as the mean of mitigation delay times computed over all the false triggers. If a false trigger is not rejected even after processing the whole utterance, we set $t_d=$\textit{utterance-length} for that utterance. We analyze the performance of our streaming FTM model by comparing $\tilde{t_d}$ versus TPR as we move the mitigation threshold. We expect our devices to have minimal $\tilde{t_d}$ at high TPR so that the false triggers are rejected as soon as possible with minimal processing of the undirected audio whereas the true triggers are rarely rejected ensuring good user experience. 

\vspace{-1mm}
\subsection{Results and Analysis}
\label{sec:results}
\vspace{-1mm}
We compare the knowledge transfer-based streaming FTM approach to two baseline systems here: 1) the teacher LatticeGNN model and 2) a LSTM-based acoustic FTM model which is not trained using the knowledge transfer framework ($\alpha=0.0$). LatticeGNN FTM system acts as a strong baseline but it has more parameters and uses more computational resources than LSTM based acoustic FTM models. On the other hand, the baseline acoustic FTM has the same architecture and number of parameters as the knowledge transfer-based student acoustic FTM. But it is expected to be a weak baseline as it does not leverage any information from LatticeGNN teacher model. During the training of student acoustic FTM, $\alpha=0.1$ was found to be optimal for best mitigation performance on our \textit{CV} set. Both the baseline and student acoustic FTM are trained and optimized using the same data. Note that the LatticeGNN model consumes ASR lattices obtained by processing entire utterances only and therefore, it is not a streaming model.  Also, ASR lattices of abruptly clipped utterances (i.e. utterance not clipped at word or sentence boundaries) are not of good quality for training an accurate FTM model. The performance of LatticeGNN model should only be compared to the streaming acoustic FTM models when $t_d=$\textit{utt-length}.

\begin{table}[b]
\small
\centering
\begin{tabular}{c|ccc|ccc}
\hline
\multirow{2}{*}{$t_d$ (in s)}&\multicolumn{3}{c|}{AUC} & \multicolumn{3}{c}{FAR (at TPR=0.99)}\\\cline{2-7}
&BL&ST&LGNN&BL&ST&LGNN\\
\hline
1.0 & 0.960 & 0.978 & - & 0.44 & 0.31 & -\\ 
1.5 & 0.970 & 0.985 & - & 0.36 & 0.24 & -\\ 
2.0 & 0.974 & 0.986 & - & 0.34 & 0.21 & -\\ 
2.5 & 0.976 & 0.988 & - & 0.32 & 0.19 & -\\
utt-len & 0.985 & \textbf{0.991} & 0.990 & 0.20 & \textbf{0.13} & 0.16\\
\hline
\end{tabular}
\caption{AUC and FAR for plots in Figure \ref{fig:fixed_mdt}. BL=Baseline AcousticFTM, ST=Student AcousticFTM, LGNN=LatticeGNN.}
\label{tab:results}
\end{table}

Figure \ref{fig:fixed_mdt} (a) and (b) show ROC curves using the fixed MDT approach detailed in Section \ref{sec:fmdt} for different values of $t_d$. As expected, the FTM accuracy improves monotonically with increasing $t_d$ for both baseline and student acoustic FTM models. For all values of $t_d$, the student acoustic FTM outperforms the baseline acoustic FTM demonstrating the efficacy of our knowledge transfer approach.  For $t_d=$\textit{utt-length}, the student acoustic FTM further outperforms the teacher LatticeGNN model as well. Table \ref{tab:results} summarizes these results in terms of AUC and FAR metrics. The student acoustic FTM has the lowest FAR at TPR=0.99 and highest AUC  metric. After processing only 2s of the utterances, the student acoustic FTM can successfully mitigate $\sim80\%$ of false triggers while rejecting only $\sim1\%$ of the true triggers.

\begin{figure}[htp]
  \centering
  \includegraphics[width=\linewidth]{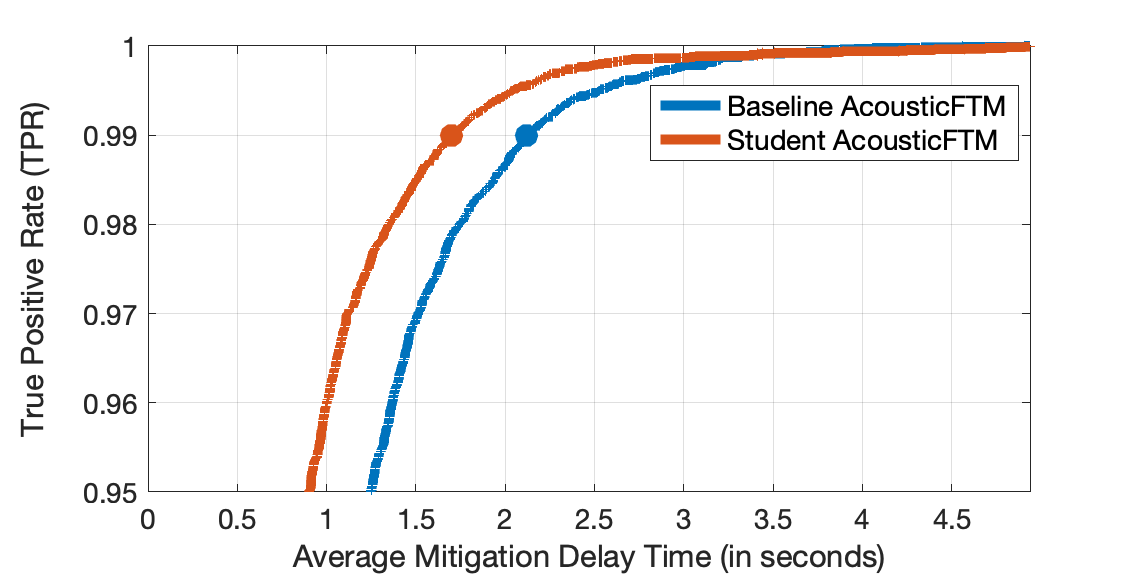}
  \caption{Streaming FTM with variable MDT.}
  \label{fig:variable_mdt}
\end{figure}

Figure \ref{fig:variable_mdt} shows a plot between average mitigation delay time $\tilde{t_d}$ and TPR under the streaming FTM approach explained in Section \ref{sec:vmdt}. We observe that for any chosen TPR, the student acoustic FTM has smaller value of $\tilde{t_d}$ than the baseline acoustic FTM which means that we need to process less unintended audio to make the correct mitigation decisions. For example, at TPR=0.99, the student acoustic FTM requires only $\tilde{t_d}=1.69$s as compared to $\tilde{t_d}=2.12$s for baseline acoustic FTM. The improved performance of student acoustic FTM indicates that the knowledge transfer approach succeeds in transferring utterance-level knowledge from lattice embeddings to the framewise acoustic embeddings of the LSTM networks. Now, the acoustic embeddings from student network possess more accurate information about the device-directedness of the utterance from early on in the utterance than the embeddings from the baseline acoustic FTM, thereby resulting in early and accurate false trigger mitigation. Furthermore, the average duration of false triggers in \textit{Eval} set is $\sim5.44$s which suggests that we can reject majority of false triggers much earlier without the need of running ASR on the complete utterances as done in the LatticeGNN approach.

\section{Conclusions}
\label{sec:conclusions}
\vspace{-2mm}
In this work, we explored a streaming false trigger mitigation approach to minimize false triggers on voice-enabled smart assistant devices. The proposed approach can stop the processing of unintended audio as soon as there is enough evidence accumulated that the audio is not directed towards the device. Moreover, it does not need to transcribe the spoken content using ASR - which not only preserves user privacy but also alleviates the need of heavy computational resources. Using knowledge-transfer technique, we demonstrated that the performance of our small-footprint acoustic features based FTM can outperform the ASR-based FTM techniques. In future, the streaming FTM training can be modified such that the network is optimized to penalize late decisions and favor early mitigation of false triggers without rejecting true triggers. The proposed approach could also be extended to other tasks such as on-device streaming user intent classification.

\vspace{-2mm}
{\footnotesize
\section{Acknowledgments}
\vspace{-2mm}
We would like to thank our colleagues John Bridle and Russ Webb for their insightful comments on this work.}
%The ISCA Board would like to thank the organizing committees of the past INTERSPEECH conferences for their help and for kindly providing the template files. \\
%Note to authors: Authors should not use logos in acknowledgement section; rather authors should acknowledge corporations by naming them only.

\bibliographystyle{IEEEbib}
\newpage
\bibliography{mybib}

\end{document}